\documentclass[11pt]{article}

\usepackage{graphicx}
\usepackage{psfrag}
\usepackage[small]{caption2}
\usepackage[below]{placeins}
\usepackage{flafter}
\usepackage{amssymb}
\usepackage{amsmath}

\newcommand{\non}{\nonumber}
\def\be{\begin{equation}}
\def\ee{\end{equation}}
\def\bestar{\begin{equation*}}
\def\eestar{\end{equation*}}
\def\bea{\begin{eqnarray}}
\def\eea{\end{eqnarray}}
\def\beastar{\begin{eqnarray*}}
\def\eeastar{\end{eqnarray*}}

\def\sket#1{\left| #1\right\rangle}           
\def\sVEV#1{\left\langle #1\right\rangle}     

\def\a{\alpha}
\def\b{\beta}
\def\d{\delta}

\def\j{\psi}

\def\p{\pi}           
\def\s{\sigma}        

\def\cp{{\cal P}}

\begin{document}

\title{{\bf Shor's Algorithm for Factoring Large Integers\footnote{Contents based
on lecture notes from graduate courses in Quantum Computation given at LNCC.}}}

\author{C. Lavor$^\dagger$, L.R.U. Manssur$^{\ddagger}$,
and R. Portugal$^{\ddagger}$\\
\\
{\mbox{}$^\dagger$}{\small Instituto de Matem\'{a}tica e
Estat\'{\i}stica }\\ {\small Universidade do Estado do Rio de
Janeiro - UERJ}\\ {\small Rua
S\~{a}o Francisco Xavier, 524, 6${{}^o}$%
andar, bl. D, sala 6018,}\\ {\small Rio de Janeiro, RJ, 20550-900,
Brazil}\\ \textit{e-mail: carlile@ime.uerj.br} \and
\\
\mbox{}{$^{\ddagger}$}{\small Coordena\c{c}\~{a}o de Ci\^encia da
Computa\c{c}\~{a}o}\\ {\small Laborat\'{o}rio Nacional de
Computa\c{c}\~{a}o Cient\'{\i}fica - LNCC}\\ {\small Av.
Get\'{u}lio Vargas 333, Petr\'{o}polis, RJ, 25651-070, Brazil}\\
\textit{e-mail: \{leon,portugal\}@lncc.br} } \maketitle

\begin{abstract}

{\normalsize \noindent This work is a tutorial on Shor's factoring
algorithm by means of a worked out example. Some basic concepts of
Quantum Mechanics and quantum circuits are reviewed. It is
intended for non-specialists which have basic knowledge on
undergraduate Linear Algebra. }

\end{abstract}

\section{Introduction}

In the last 30 years, the number of transistors per chip roughly
doubled every 18 months, amounting to an exponentially growing
power of classical computers. Eventually this statement (Moore's
law) will be violated, since the transistor size will reach the
limiting size of one atom in about 15 years. Even before that,
disturbing quantum effects will appear.

Ordinarily one states that if an algorithm is inefficient, one
simply waits for hardware efficient enough to run it. If the
exponential increase in the power of classical computers becomes
saturated, the class of inefficient algorithms will remain
useless. From this pessimistic point of view, Computer Science
seems to face very narrow limitations in the near future, coming
from the physics underneath computer architecture.

It is important to keep in mind that a computer is a device
governed by the laws of Physics. For decades, this fact was
irrelevant. Computer Science emerged in a mathematical context and
the specifics imposed by Physics were so few that most computer
scientists paid no attention to them. One possible explanation for
this state of matters is that computers work under the laws of
classical physics, which are common sense.

In a seminal paper, Feynman \cite{feynman} argued that the way
classical computers work is a special case of some more general
form allowed by the laws of Quantum Mechanics. He gave general
arguments supporting the idea that a manifestly quantum device
would be exponentially faster than a classical one. Subsequently,
Deutsch \cite{deutsch} generalized the classical circuit model to
its quantum counterpart and gave the first example of a quantum
algorithm faster than its classical counterpart. Based on
Deutsch's work, Simon \cite{simon} developed a quantum algorithm
exponentially faster than its classical counterpart, taking
advantage of entanglement, corroborating with Feynman's arguments.

The greatest success came with Shor's work \cite{shor}. He
developed exponentially faster quantum algorithms for factoring
integers and for finding discrete logarithms when compared to the
known classical algorithms. Shor's algorithms allow one to render
most current cryptographic methods useless, when a quantum
computer of reasonable size is available.

This work is an introductory review of Shor's factoring algorithm.
We have put all our efforts to write as clear as possible for
non-specialists. We assume familiarity with undergraduate Linear
Algebra, which is the main mathematical basis of Quantum
Mechanics. Some previous knowledge of Quantum Mechanics is
welcome, but not necessary for a persistent reader. The reader can
find further material in
\cite{shor,jozsa,chuang,preskill,aharonov}.

Section 2 reviews basic notions of Quantum Mechanics necessary for
Quantum Computation. Section 3 introduces the notion of quantum
circuits and presents some basic examples. Section 4 describes how
factorization can be reduced to order calculation and Section 5
gives a quantum algorithm for it. Section 6 shows the quantum
Fourier transform. Section 7 gives an example and finally Section
8 shows the decomposition of the Fourier transform circuit in
terms of the universal gates.

\section{Review of Quantum Mechanics for Quantum Computation}

\begin{figure}
    \setcaptionmargin{.5in}
    \centering
    \psfrag{1}{\footnotesize $\sket 1$}
    \psfrag{psi}{\footnotesize $\sket \psi$}
    \psfrag{0}{\footnotesize $\sket 0$}
    \includegraphics[height=4.5cm]{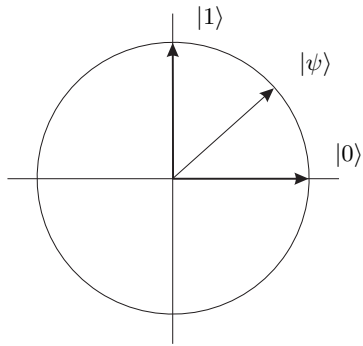}
    \caption[text for list of figures]{Computational basis for the case
$\a$, $\b$ real. In the general case ($\a$, $\b$ complex) there is
still a geometrical representation called the Bloch sphere (see
\cite{chuang} page 15).}
    \label{fig05}
\end{figure}

In classical computers, a bit can assume only values 0 or 1. In
quantum computers, the values 0 and 1 are replaced by the vectors
$|0\rangle$ and $|1\rangle$. This notation for vectors is called
the Dirac notation  and is standard in Quantum Mechanics. The name
bit is replaced by \textit{qubit}, short of \textit{quantum bit}.
The difference between bits and qubits is that a qubit $|\psi
\rangle $ can also be in a linear combination of the vectors
$|0\rangle $ and $|1\rangle $,
\begin{equation}
|\psi \rangle =\alpha |0\rangle +\beta |1\rangle ,  \label{a}
\end{equation}
where $\alpha $ and $ \beta $ are complex numbers. $|\psi \rangle$
is said to be a {\em superposition} of the vectors $|0\rangle $
and $|1\rangle $ with {\em amplitudes} $\alpha $ and $\beta $.
Thus, $|\psi \rangle $ is a vector in a two-dimensional complex
vector space, where \{$|0\rangle $, $|1\rangle\} $ forms an
orthonormal basis, called the {\it computational basis} (see Fig.
\ref{fig05} in the real case). The state $|0\rangle $ is not the
zero vector, but simply the first vector of the basis. The matrix
representations of the vectors $|0\rangle $ and $|1\rangle $ are
given by
\begin{equation*}
|0\rangle =\left[
\begin{array}{c}
1 \\ 0
\end{array}
\right] \;\;\;\mbox{ and }\;\;\; |1\rangle =\left[
\begin{array}{c}
0 \\ 1
\end{array}
\right] .
\end{equation*}
In Quantum Mechanics, vectors are systematically called
\textit{states}. We use this term from now on.

The physical interpretation of $|\psi \rangle $ is that it
coexists in two states: $|0\rangle $ and $|1\rangle $. It is
similar to a coin that is partially heads up and partially tails
up simultaneously. We cannot push further the analogy simply
because quantum phenomena do not have a classical analogue in
general. The state $|\psi \rangle $ can store a huge quantity of
information in its coefficients $\alpha$ and $\beta$, but this
information lives in the quantum level, which is microscopic
(usually quantum effects appear in atomic dimensions). To bring
quantum information to the classical level, one must measure the
qubit. Quantum Mechanics tells us that the measurement process
inevitably disturbs a qubit state, producing a non-deterministic
collapse of $|\psi \rangle $ to either $|0\rangle $ or $|1\rangle
$. One gets $|0\rangle $ with probability  $|\alpha |^{2}$ or
$|1\rangle$ with probability $|\beta |^{2}$. The non-deterministic
collapse does not allow one to determine the values of $\alpha$
and $\beta$ before the measurement.  They are inaccessible via
measurements unless one has many copies of the same state. Two
successive measurements of the same qubit give the same output. If
$|\alpha |^{2}$ and $|\beta |^{2}$ are probabilities and there are
only two possible outputs, then
\begin{equation}
|\alpha |^{2}+|\beta |^{2}=1.  \label{b}
\end{equation}
Calculating the norm of $|\psi \rangle $, Eq. (\ref{b}) gives
\begin{equation*}
||\;|\psi \rangle \;||=\sqrt{|\alpha |^{2}+|\beta |^{2}} = 1.
\end{equation*}

A measurement is not the only way that one can interact with a
qubit. If one does not obtain any information about the state of
the qubit, the interaction changes the values of $\alpha $ and
$\beta $ keeping the constraint (\ref{b}). The most general
transformation of this kind is a linear transformation $U$ that
takes unit vectors into unit vectors. Such a transformation is
called \textit{unitary} and can be defined by
\begin{equation*}
U^{\dagger }U=UU^{\dagger }=I,
\end{equation*}
where $U^{\dagger }$ $=(U^{\ast })^{T}$ ($\ast $ indicates complex
conjugation and $T$ indicates the transpose operation) and $I$ is
the $2\times 2$ identity matrix.

So far we are dealing with one-qubit quantum computers. To
consider the multiple qubit case, it is necessary to introduce the
concept of \textit{tensor product}. Suppose $V$ and $W$ are
complex vector spaces of
dimensions $m$ and $n$, respectively. The tensor product $V\otimes W$ is an $%
mn$-dimensional vector space. The elements of $V\otimes W$ are
linear combinations of tensor products $|v\rangle \otimes
|w\rangle $, satisfying the following properties ($z\in
\mathbb{C}$, $|v\rangle ,|v_{1}\rangle ,|v_{2}\rangle \in V$, and
$|w\rangle ,|w_{1}\rangle ,|w_{2}\rangle \in W$):

\begin{enumerate}
\item  $z(|v\rangle \otimes |w\rangle )=(z|v\rangle )\otimes |w\rangle
=|v\rangle \otimes (z|w\rangle ),$

\item  $(|v_{1}\rangle +|v_{2}\rangle )\otimes |w\rangle =(|v_{1}\rangle
\otimes |w\rangle )+(|v_{2}\rangle \otimes |w\rangle ),$

\item  $|v\rangle \otimes (|w_{1}\rangle +|w_{2}\rangle )=(|v\rangle \otimes
|w_{1}\rangle )+(|v\rangle \otimes |w_{2}\rangle ).$
\end{enumerate}

\noindent We use also the notations $|v\rangle |w\rangle ,$
$|v,w\rangle $ or
$%
|vw\rangle $ for the tensor product $|v\rangle \otimes |w\rangle
$. Note that the tensor product is non-commutative, so the
notation must preserve the ordering.

Given two linear operators $A$\ and $B$\ defined on the vector
spaces $V$\
and $W$, respectively, we can define the linear operator $A\otimes B$\ on $%
V\otimes W$\ as
\begin{equation}
(A\otimes B)(|v\rangle \otimes |w\rangle )=A|v\rangle \otimes
B|w\rangle,\label{c}
\end{equation}
where $|v\rangle \in V$ and $|w\rangle \in W$. The matrix
representation of $A\otimes B$ is given by
\begin{equation}
A\otimes B=\left[
\begin{array}{rrr}
A_{11}B & \cdot \cdot \cdot & A_{1m}B \\ \vdots & \ddots & \vdots
\\ A_{m1}B & \cdot \cdot \cdot & A_{mm}B
\end{array}
\right] ,  \label{c1}
\end{equation}
where $A$ is an $m\times m$ matrix and $B$ is a $n\times n$ matrix
(we are using the same notation for the operator and its matrix
representation). So, the matrix $A\otimes B$ has dimension
$mn\times mn$. For example, given
\begin{equation*}
A=\left[
\begin{array}{rr}
0 & 1 \\ 1 & 0
\end{array}
\right] \;\;\;\text{  and  }\;\;\;B=\left[
\begin{array}{rrr}
1 & 0 & 0 \\ 0 & 1 & 0 \\ 0 & 0 & 1
\end{array}
\right] ,
\end{equation*}
the tensor product $A\otimes B$ is
\begin{equation*}
A\otimes B=\left[
\begin{array}{rr}
0 & 1 \\ 1 & 0
\end{array}
\right] \otimes \left[
\begin{array}{rrr}
1 & 0 & 0 \\ 0 & 1 & 0 \\ 0 & 0 & 1
\end{array}
\right] =\left[
\begin{array}{rrrrrr}
0 & 0 & 0 & 1 & 0 & 0 \\ 0 & 0 & 0 & 0 & 1 & 0 \\ 0 & 0 & 0 & 0 &
0 & 1 \\ 1 & 0 & 0 & 0 & 0 & 0 \\ 0 & 1 & 0 & 0 & 0 & 0 \\ 0 & 0 &
1 & 0 & 0 & 0
\end{array}
\right] .
\end{equation*}
The formula (\ref{c1}) can also be used for non-square matrices,
such as the tensor product of two vectors. For example, if we have
a 2-qubit quantum computer and the first qubit is in the state
$|0\rangle$ and the second is in the state $|1\rangle$, then the
quantum computer is in the state $|0\rangle \otimes |1\rangle$,
given by
\begin{equation}
|0\rangle \otimes |1\rangle =|01\rangle =\left[
\begin{array}{r}
1 \\ 0
\end{array}
\right] \otimes \left[
\begin{array}{r}
0 \\ 1
\end{array}
\right] =\left[
\begin{array}{r}
0 \\ 1 \\ 0 \\ 0
\end{array}
\right] . \label{ket01}
\end{equation}
The resulting vector is in a 4-dimensional vector space.


The general state $|\psi \rangle $ of a 2-qubit quantum computer
is a superposition of the states $|00\rangle ,$ $|01\rangle $,
$|10\rangle $, and $|11\rangle $,
\begin{equation}
|\psi \rangle =\alpha |00\rangle +\beta |01\rangle +\gamma
|10\rangle +\delta |11\rangle ,  \label{d}
\end{equation}
with the constraint
\begin{equation*}
|\alpha |^{2}+|\beta |^{2}+|\gamma |^{2}+|\delta |^{2}=1.
\end{equation*}
Regarding the zeroes and ones as constituting the binary expansion
of an integer, we can replace the representations of states
\begin{equation*}
|00\rangle ,\text{ }|01\rangle ,\text{ }|10\rangle ,\text{
}|11\rangle ,
\end{equation*}
by the shorter forms
\begin{equation*}
|0\rangle ,\text{ }|1\rangle ,\text{ }|2\rangle ,\text{
}|3\rangle,
\end{equation*}
in decimal notation, which is handy in some formulas.

In general, the state $|\psi \rangle $ of an $n$-qubit quantum
computer is a
superposition of the $2^{n}$ states $|0\rangle ,$ $|1\rangle ,$ $...,$ $%
|2^{n}-1\rangle $,
\begin{equation*}
|\psi \rangle =\underset{i=0}{\overset{2^{n}-1}{\sum }}\alpha
_{i}|i\rangle ,
\end{equation*}
with amplitudes $\alpha _{i}$ constrained to
\begin{equation*}
\underset{i=0}{\overset{2^{n}-1}{\sum }}|\alpha _{i}|^{2}=1.
\end{equation*}
Recall that the orthonormal basis $\{ \sket{0}, \dots,
\sket{2^n-1} \}$ is the computational basis in decimal notation.
The state of an $n$-qubit quantum computer is a vector in a
$2^n$-dimensional complex vector space. When the number of qubits
increases linearly, the dimension of the associated vector space
increases exponentially. As before, a measurement of a generic
state $\sket \psi$ yields the result $\sket{i_0}$ with probability
$|\alpha _{i_0}|^{2}$, where $0 \leq i_0 < 2^{n}$. Usually, the
measurement is performed qubit by qubit yielding zeroes or ones
that are read together to form $i_0$. We stress again a very
important feature of the measurement process. The state $\sket
\psi$ as it is before measurement is inaccessible unless it is in
the computational basis. The measurement process inevitably
disturbs $\sket \psi$ forcing it to collapse to one vector of the
computational basis. This collapse is non-deterministic, with the
probabilities given by the squared norms of the corresponding
amplitudes in $\sket \psi$.

If we have a 2-qubit quantum computer, the first qubit in the
state
\begin{equation*}
|\varphi \rangle =a|0\rangle +b|1\rangle
\end{equation*}
and the second in the state
\begin{equation*}
|\psi \rangle =c|0\rangle +d|1\rangle ,
\end{equation*}
then the state of the quantum computer is the tensor product
\begin{eqnarray}
|\varphi \rangle \otimes |\psi \rangle &=&(a|0\rangle +b|1\rangle
)\otimes (c|0\rangle +d|1\rangle )  \label{e} \\ &=&ac|00\rangle
+ad|01\rangle +\ bc|10\rangle +bd|11\rangle . \notag
\end{eqnarray}
Note that a general $2$-qubit state (\ref{d}) is of the form
(\ref{e}) if and only if
\begin{eqnarray*}
\alpha &=&ac, \\ \beta &=&ad, \\ \gamma &=&bc, \\ \delta &=&bd.
\end{eqnarray*}
From these equalities we have that a general $2$-qubit state
(\ref{d}) is of the form (\ref{e}) if and only if
\begin{equation*}
\alpha \delta =\beta \gamma .
\end{equation*}
Thus, the general $2$-qubit state is not necessarily a product of
two one-qubit states. Such non-product states of two or more
qubits are called \textit{entangled} states, for example,
$(\sket{00} + \sket{11}) / \sqrt 2$. The entangled states play an
essential role in quantum computation. Quantum computers that do
not use entanglement cannot be exponentially faster than classical
computers. On the other hand, a naive use of entanglement does not
guarantee any improvements.

A complex vector space $V$ is a Hilbert space if there is an
\textit{inner product}, written in the form $\langle \varphi |\psi
\rangle $, defined by the following rules ($a,b\in \mathbb{C}$ and
$|\varphi \rangle ,|\psi \rangle ,|u\rangle ,|v\rangle \in V$):

\begin{enumerate}
\item  $\langle \psi |\varphi \rangle =\langle \varphi |\psi \rangle ^{\ast
},$

\item  $\langle \varphi |(a|u\rangle +b|v\rangle )\rangle =a\langle \varphi
|u\rangle +b\langle \varphi |v\rangle ,$

\item  $\langle \varphi |\varphi \rangle >0$ if $|\varphi \rangle \neq 0.$
\end{enumerate}

\noindent The {\em norm} of a vector $|\varphi \rangle $ is given
by
\begin{equation*}
||\;|\varphi \rangle \;||=\sqrt{\langle \varphi |\varphi \rangle
}.
\end{equation*}

The notation $\langle \varphi |$ is used for the \textit{%
dual vector} to the vector $|\varphi \rangle $. The dual is a
linear operator from the vector space $V$ to the complex numbers,
defined by
\begin{equation*}
\langle \varphi |(|v\rangle )=\langle \varphi |v\rangle , \;\;\;
\text{   } \forall |v\rangle \in V.
\end{equation*}

Given two vectors $|\varphi \rangle $ and $|\psi \rangle $ in a
vector space $V$, there is also an \textit{outer product} $|\psi
\rangle \langle \varphi | $, defined as a linear operator on $V$
satisfying
\begin{equation*}
(|\psi \rangle \langle \varphi |)|v\rangle =|\psi \rangle \langle
\varphi |v\rangle , \;\;\; \text{   }\forall |v\rangle \in V.
\end{equation*}

If $|\varphi \rangle =a |0\rangle +b |1\rangle $ and $|\psi
\rangle =c |0\rangle +d |1\rangle $, then the matrix
representations for inner and outer products are:
\begin{eqnarray*}
\langle \varphi |\psi \rangle &=&\left[
\begin{array}{rr}
a ^{\ast } & b ^{\ast }
\end{array}
\right] \left[
\begin{array}{r}
c \\ d
\end{array}
\right] =a ^{\ast }c +b ^{\ast }d, \\ |\varphi \rangle \langle
\psi | &=&\left[
\begin{array}{r}
a \\ b
\end{array}
\right] \left[
\begin{array}{rr}
c ^{\ast } & d ^{\ast }
\end{array}
\right] =\left[
\begin{array}{rr}
ac ^{\ast } & ad ^{\ast } \\ bc ^{\ast } & bd ^{\ast }
\end{array}
\right] .
\end{eqnarray*}
The matrix of the outer product is obtained by usual matrix
multiplication of a column matrix by a row matrix. But in this
case, we can replace the matrix multiplication by the tensor
product, i.e., $|\varphi \rangle \langle \psi |=|\varphi \rangle
\otimes \langle \psi |$ (notice the complex conjugation in the
process of taking the dual).
\begin{figure}
    \setcaptionmargin{.5in}
    \centering
    \psfrag{bit0}{\footnotesize $\sket {\psi_1}$}
    \psfrag{bit1}{\footnotesize $\sket {\psi_2}$}
    \psfrag{bitn-1}{\footnotesize$\sket {\psi_n}$}
    \psfrag{bit0p}{\footnotesize$\sket {\psi^\prime_1}$}
    \psfrag{bit1p}{\footnotesize$\sket {\psi^\prime_2}$}
    \psfrag{bitn-1p}{\footnotesize$\sket {\psi^\prime_n}$}
    \psfrag{psi}[][]{\footnotesize$\sket {\psi}$}
    \psfrag{f}{$U$}
    \includegraphics[]{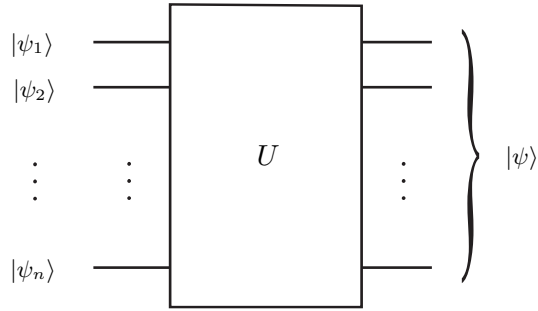}
    \caption[text for list of figures]{The sketch of the quantum computer. We
    consider the input nonentangled, which is reasonable in general. On the
other hand,
    the output is entangled in general. The measurement of the state $\sket
\psi$,
    not shown here, returns zeroes and ones.}
    \label{fig06}
\end{figure}

After the above review, we are ready to outline the quantum
computer. In Fig. \ref{fig06}, we are taking a nonentangled input,
what is quite reasonable. In fact, $\sket{\psi_i}$ is either
$\sket 0$ or $\sket 1$ generally. $\sket \psi$, on the right hand
side of Fig. \ref{fig06}, is the result of the application of a
unitary operator $U$ on the input. The last step is the
measurement of the states of each qubit, which returns zeroes and
ones that form the final result of the quantum calculation. Note
that there is, in principle, an infinite number of possible
operators $U$, which are unitary $2^n\times 2^n$ matrices.


\section{Quantum Circuits}

%
\begin{figure}
    \setcaptionmargin{.5in}
    \centering
    \vspace{.3in}
    \psfrag{psi}{\footnotesize$\sket \psi$}
    \psfrag{Xpsi}{\footnotesize$X \sket \psi$}
    \includegraphics{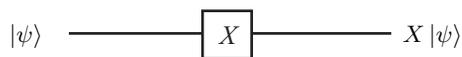}
    \caption[text for list of figures]{Quantum NOT gate.}
    \label{fig07}
\end{figure}
Let us start with one-qubit gates. In the classical case there is
only one possibility, which is the NOT gate. The NOT gate inverts
the bit value: 0 goes to 1 and vice-versa. The straightforward
generalization to the quantum case is given in Fig. \ref{fig07},
where $X$ is the unitary operator
\begin{equation*}
X=\left[
\begin{array}{cc}
0 & 1 \\ 1 & 0
\end{array}
\right] .
\end{equation*}
So, if the input $|\psi \rangle $ is $|0\rangle$, the output is
$|1\rangle $ and vice-versa. But now we can have a situation with
no classical counterpart. The
state $|\psi \rangle $ can be a superposition of states $|0\rangle $ and $%
|1\rangle $. The general case is $|\psi \rangle = \alpha |0\rangle
+ \beta |1\rangle $ and the corresponding output is $\alpha
|1\rangle +\beta |0\rangle $.

The gate $X$ is not the only one-qubit gate. There are infinitely
many, since there are an infinite number of $2\times 2$ unitary
matrices. In principle, any unitary operation can be implemented
in practice. The {\em Hadamard} gate is another important
one-qubit gate, given by
\begin{equation*}
H=\frac{1}{\sqrt{2}}\left[
\begin{array}{rr}
1 & 1 \\ 1 & -1
\end{array}
\right] .
\end{equation*}
It is easy to see that
\begin{eqnarray*}
H|0\rangle  &=&\frac{|0\rangle +|1\rangle }{\sqrt{2}}, \\
H|1\rangle  &=&\frac{|0\rangle -|1\rangle }{\sqrt{2}}.
\end{eqnarray*}
If the input is $|0\rangle $, the Hadamard gate creates a
superposition of states with equal weights. This is a general
feature, valid for two or more qubits. Let us analyze the
$2$-qubit case.

The first example of a $2$-qubit gate is $H\otimes H$:
\begin{eqnarray*}
H^{\otimes 2}|0\rangle |0\rangle  &=&(H\otimes H)(|0\rangle
\otimes |0\rangle )=H|0\rangle \otimes H|0\rangle  \\ &=&\left(
\frac{|0\rangle +|1\rangle }{\sqrt{2}}\right) \otimes \left(
\frac{%
|0\rangle +|1\rangle }{\sqrt{2}}\right)  \\
&=&\frac{1}{2}(|0\rangle |0\rangle +|0\rangle |1\rangle +|1\rangle
|0\rangle +|1\rangle |1\rangle ) \\ &=&\frac{1}{2}(|0\rangle
+|1\rangle +|2\rangle +|3\rangle ).
\end{eqnarray*}
The result is a superposition of all basis states with equal
weights. More generally, the Hadamard operator applied to the
$n$-qubit state $|0\rangle$ is \bea H^{\otimes n}|0\rangle & = &
H^{\otimes n} \sket{0,...,0}  = (H \sket{0})^{\otimes n} \non
\\
 & = & \left(\frac{\sket{0} + \sket{1}}{\sqrt{2}}\right)^{\otimes n}
\non \\
 & = & \frac{1}{\sqrt{2^{n}}} \sum_{i=0}^{2^{n}-1} \sket{i} .\label{psi00}
\eea
Thus, the tensor product of $n$ Hadamard operators produces an
equally weighted \ superposition of all computational basis
states, when the input is the state $|0\rangle.$ This state is
useful for applying quantum parallelism, as we will see ahead.

Another important $2$-qubit quantum gate is the CNOT gate. It has
two input qubits, the control and the target qubit, respectively.
The target qubit is flipped only if the control qubit is set to 1,
that is,
\bea |00\rangle & \rightarrow & |00\rangle, \non \\ |01\rangle &
\rightarrow & |01\rangle, \label{cnot}\\ |10\rangle & \rightarrow
& |11\rangle, \non \\ |11\rangle & \rightarrow & |10\rangle . \non
\eea
The action of the CNOT gate can also be represented by
\begin{equation*}
|a,b\rangle \rightarrow |a,a\oplus b\rangle ,
\end{equation*}
\begin{figure}
    \setcaptionmargin{.5in}
    \centering
    \psfrag{i}{\footnotesize$\sket {i}$}
    \psfrag{s}{\footnotesize$\sket {\s}$}
    \psfrag{xisigma}{\footnotesize$X^i \sket {\s}$}
    \includegraphics{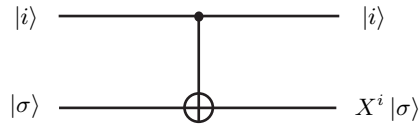}
    \caption{CNOT gate. $|i\rangle$ can be either $|0\rangle$ or
     $|1\rangle$. The general case is obtained by linearity. }
    \label{fig016}
\end{figure}
where $\oplus $ is addition modulo 2. Now, let us obtain its
matrix representation. Performing the same calculations that yield
Eq. (\ref{ket01}), we have
\begin{equation}
|00\rangle  = \left[
\begin{array}{r}
1 \\ 0 \\ 0 \\ 0
\end{array}
\right],\,\,\ \  |01\rangle  =\left[
\begin{array}{r}
0 \\ 1 \\ 0 \\ 0
\end{array}
\right],\,\,\ \  |10\rangle   =\left[
\begin{array}{r}
0 \\ 0 \\ 1 \\ 0
\end{array}
\right],\,\,\ \  |11\rangle  =\left[
\begin{array}{r}
0 \\ 0 \\ 0 \\ 1
\end{array}
\right].  \label{vectors}
\end{equation}
Thus, from (\ref{cnot}) and (\ref{vectors}), the matrix
representation $U_{\mbox{\tiny CNOT}}$ of the CNOT gate is
\begin{equation*}
U_{\mbox{\tiny CNOT}}=\left[
\begin{array}{rrrr}
1 & 0 & 0 & 0 \\ 0 & 1 & 0 & 0 \\ 0 & 0 & 0 & 1 \\ 0 & 0 & 1 & 0
\end{array}
\right] . \label{Ucnot}
\end{equation*}

Fig. \ref{fig016} describes the CNOT gate, where $\sket i$ is
either $\sket 0$ or $\sket 1$. The figure could lead one to think
that the output is always nonentangled, but that is not true,
since if the first qubit is in a more general state given by
$a\sket 0 + b \sket 1$, then the output will be $a\sket 0 \sket
\s+ b \sket 1 X \sket \s$, which is entangled in general. The
input can be entangled as well.

We have seen two examples of $2$-qubit gates. The general case is
a $4\times 4$ unitary matrix. Gates that are the direct product of
other gates, such as $H\otimes H$, do not produce entanglement. If
the input is nonentangled, the output is not too. On the other
hand, the output of the CNOT gate can be entangled while the input
is nonentangled.

CNOT and one-qubit gates form a universal set of gates. This means
that any other gate, operating on $2$ or more qubits, can be
written as compositions and direct products of CNOT and one-qubit
gates \cite{chuang}.

At the end of Section 2, we gave a general outline of the quantum
computer (Fig. 2) based on the action of a unitary operator $U$.
In the present section, we have seen that in general $U$ can be
broken up in terms of smaller gates. This decomposition is useful
because it corresponds to the natural steps that describe an
algorithm. So, a quantum algorithm consists of a sequence of
unitary operators acting on sets of qubits. These unitary
operators multiplied together form the operator $U$ of Fig. 2.

More details on quantum circuits can be found in \cite{chuang,
barenco}.

\section{Factorization can be reduced to order calculation}

Let us describe Shor's algorithm for finding the prime factors of
a composite number $N$. Think of a large number such as one with
300 digits in decimal notation, since such numbers are used in
cryptography. Though $N$ is large, the number of qubits necessary
to store it is small. In general $\log_2 N$ is not an integer, so
let us define
\bestar n=\lceil \log_2 N \rceil. \eestar
A quantum computer with $n$ qubits can store $N$ or any other
positive integer less than $N$. With a little thought, we see that
the number of prime factors of $N$ is at most $n$. If both the
number of qubits and the number of factors are less than or equal
to $n$, then it is natural to ask if there is an algorithm that
factors $N$ in a number of steps which is polynomial in $n$. More
technically, the question is: is there a factorization algorithm
in the complexity class $\cp$ \cite{papadimitriou}?

Reduction of factorization of $N$ to the problem of finding the
\textit{order} of an integer $x$ less than $N$ is as follows. If
$x$ and $N$ have common factors, then GCD$(x,N)$ gives a factor of
$N$, therefore it suffices to investigate the case when $x$ is
coprime to $N$. The order of $x$ modulo $N$ is defined as the
least positive integer $r$ such that
$$ \label{cond} x^r \equiv 1 \mbox{ mod } N. $$
If $r$ is even, we can define $y$ by
$$x^{r/2} \equiv y  \mbox{ mod } N.$$
The above notation means that $y$ is the remainder of $x^{r/2}$
divided by $N$ and, by definition, $0\leq y < N$. Note that $y$
satisfies $y^2 \equiv 1 \mbox{ modulo } N$, or equivalently
$(y-1)(y+1) \equiv 0 \mbox{ modulo } N$, which means that $N$
divides $(y-1)(y+1)$. If $1 < y
< N-1$, the factors $y-1$ and $y+1$ satisfy $0<y-1<y+1<N$,
therefore $N$ cannot divide $y-1$ nor $y+1$ separately. The only
alternative is that both $y-1$ and $y+1$ have factors of $N$ (that
yield $N$ by multiplication). So, GCD($y-1,N$) and GCD($y+1,N$)
yield non trivial factors of $N$ (GCD stands for the greatest
common divisor). If $N$ has remaining factors, they can be
calculated applying the algorithm recursively.

Consider $N=21$ as an example. The sequence of equivalences
\bea 2^4 & \equiv & 16 \mbox{ mod } 21 \non \\ 2^5 & \equiv & 11
\mbox{ mod } 21 \non \\ 2^6 & \equiv & 11 \times 2 \; \equiv  \; 1
\mbox{ mod }21 \non \eea
show that the order of 2 modulo 21 is $r=6$. Therefore, $y \equiv
2^3 \equiv 8 \mbox{ modulo } 21 $. $y-1$ yields the factor 7 and
$y+1$ yields the factor 3 of 21.

In summary, if we pick up at random a positive integer $x$ less
than $N$ and calculate GCD$(x,N)$, either we have a factor of $N$
or we learn that $x$ is coprime to $N$. In the latter case, if $x$
satisfies the conditions (1) its order $r$ is even, and (2)
$0<y-1<y+1< N$, then GCD($y-1,N$) and GCD($y+1,N$) yield factors
of $N$. If one of the conditions is not true, we start over until
finding a proper candidate $x$. The method would not be useful if
these assumptions were too restrictive, but fortunately that is
not the case. The method sistematically fails if $N$ is a power of
some odd prime, but an alternative efficient classical algorithm
for this case is known. If $N$ is even, we can keep dividing by 2
until the result turns out to be odd. It remains to apply the
method for odd composite integers that are not a power of some
prime number. It is cumbersome to prove that the probability of
finding $x$ coprime to $N$ satisfying the conditions (1) and (2)
is high; in fact this probability is $1 - 1/2^{k-1}$, where $k$ is
the number of prime factors of $N$. In the worst case ($N$ has 2
factors), the probability is greater than or equal to $1/2$ (see
the proof in Appendix B of \cite{jozsa}).

At first sight, it seems that we have just described an efficient
algorithm to find a factor of $N$. That is not true, since it is
not known an efficient classical algorithm to calculate the order
of an integer $x$ modulo $N$. On the other hand, there is (after
Shor's work) an efficient quantum algorithm. Let us describe it.

\section{Quantum algorithm to calculate the order}
\label{sectionorder}

\begin{figure}
    \setcaptionmargin{.6in}
    \centering
    \psfrag{Vx}{$V_x$}
    \psfrag{FT}[][]{$\mbox{DFT}^\dagger$}
    \psfrag{Ze}[][]{\footnotesize$\sket 0$}
    \psfrag{firstregister}{\footnotesize first register}
    \psfrag{tqubits}{\footnotesize (t qubits)}
    \psfrag{secondregister}{\footnotesize second register}
    \psfrag{nqubits}{\footnotesize (n qubits)}
    \psfrag{psi0}[][]{\footnotesize$\sket {\psi_0}$}
    \psfrag{psi1}[][]{\footnotesize$\sket {\psi_1}$}
    \psfrag{psi2}[][]{\footnotesize$\sket {\psi_2}$}
    \psfrag{psi3}[][]{\footnotesize$\sket {\psi_3}$}
    \psfrag{psi4}[][]{\footnotesize$\sket {\psi_4}$}
    \psfrag{psi5}[][]{\footnotesize$\sket {\psi_5}$}
    \includegraphics[]{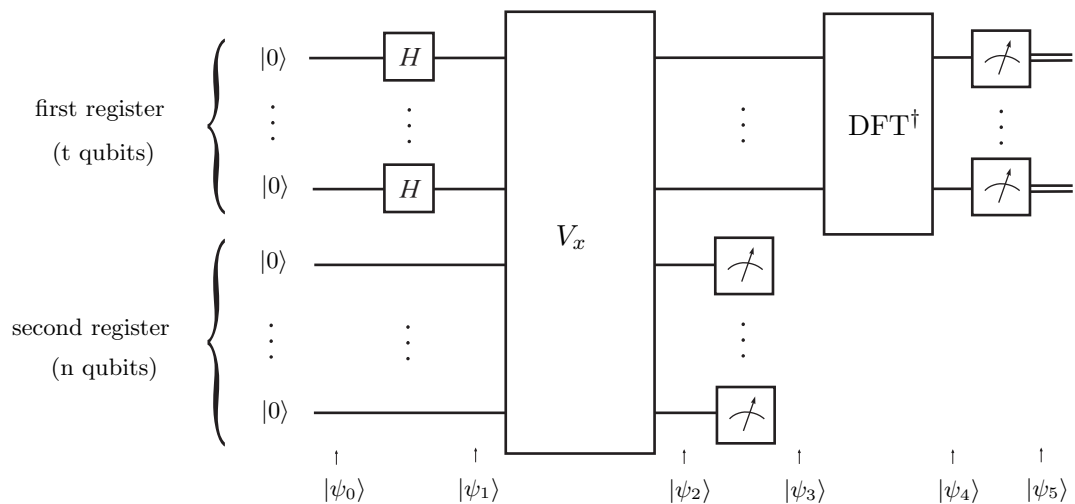}
    \caption[]{Quantum circuit for finding the order of the positive integer $x$ modulo $N$.}
    \label{order}
\end{figure}

Consider the circuit of Fig. \ref{order}. It calculates the order
$r$ of the positive integer $x$ less than $N$, coprime to $N$.
$V_x$ is the unitary linear operator
\be
\label{vx} V_x (\sket j \sket k) = \sket j \sket{k+x^j},
 \ee
where $\sket j$ and $\sket k$ are the states of the first and
second registers, respectively. The arithmetical operations are
performed modulo $N$, so $0 \leq k+x^j <N$. DFT is the Discrete
Fourier Transform operator which will be described ahead.

The first register has $t$ qubits, where $t$ is generally chosen
such that $N^2\leq 2^t < 2N^2$, for reasons that will become clear
later on \cite{shor}. As an exception, if the order $r$ is a power
of 2, then it is enough to take $t=n$. In this section we consider
this very special case and leave the general case for Section
\ref{generalization}. We will keep the variable $t$ in order to
generalize the discussion later on.

\begin{figure}
    \setcaptionmargin{.5in}
    \centering
    \psfrag{0}[][]{\footnotesize$0$}
    \psfrag{b0}[][]{\footnotesize${b_0}$}
    \psfrag{Rb0}[][]{\footnotesize${r+b_0}$}
    \psfrag{2rb0}[][]{\footnotesize${2r+b_0}$}
    \psfrag{3rb0}[][]{\footnotesize${3r+b_0}$}
    \psfrag{R2t}[][]{${\frac{r}{2^t}}$}
    \psfrag{R}[][]{\footnotesize$r$}
    \psfrag{Amplitude of each}{\footnotesize Probability distribution}
    \psfrag{terms}{\footnotesize Terms of $\sket{\psi_3}$}
    \psfrag{1st register}{\footnotesize (1st register)}
    \includegraphics[width=4.5in]{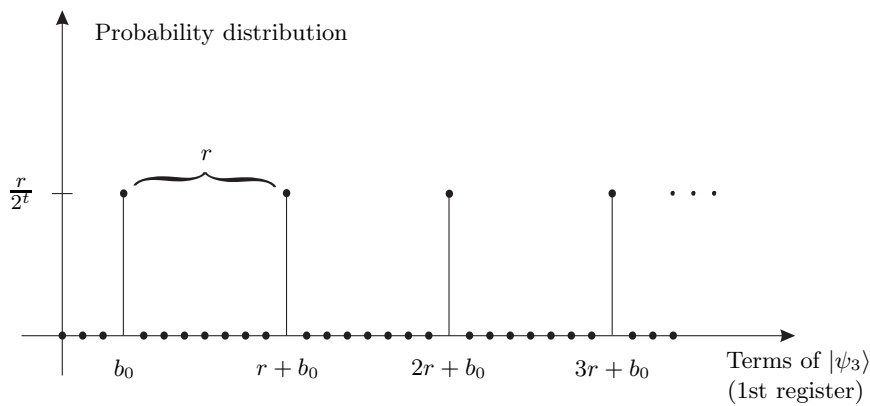}
    \caption[]{Probability distribution of $\sket {\psi_3}$
    measured in the computational basis (for the case $b_0=3$ and $r=8$).
    The horizontal axis has $2^t$ points. The number of peaks is $2^t/r$
    and the period is $r$.}
    \label{graph}
\end{figure}

The states of the quantum computer are indicated by $\sket {\j_0}$
to $\sket {\j_5}$ in Fig. \ref{order}. The initial state is
\bestar \sket{\j_0} = \underbrace {\sket{0 \ldots 0}}_t
\underbrace {\sket{0 \ldots 0}}_n. \eestar
The application of the Hadamard operator
\bestar H=\frac{1}{\sqrt{2}} \left[ \begin{array}{rr} 1 & 1 \\ 1 &
-1
\end{array} \right]
\eestar
on each qubit of the first register yields (see Eq. (\ref{psi00}))
\be \sket{\j_1} = \frac{1}{\sqrt{2^t}} \sum_{j=0}^{2^t-1} \sket j
\sket 0 . \label{psi1} \ee
The first register is then in a superposition of all states of the
computational basis with equal amplitudes given by
$\frac{1}{\sqrt{2^t}}$. Now we call the reader's attention to what
happens when we apply $V_x$ to $\sket{\j_1}$:
\bea
\sket{\j_2} & = & V_x \sket{\j_1} \non \\
 & = & \frac{1}{\sqrt{2^t}} \sum_{j=0}^{2^t-1} V_x (\sket j \sket 0) \non \\
 & = & \frac{1}{\sqrt{2^t}} \sum_{j=0}^{2^t-1} \sket j \sket{x^j}. \label{psi2}
\eea
The state $\sket{\j_2}$ is a remarkable one. Because $V_x$ is
linear, it acts on all $\sket j \sket 0 $ for $2^t$ values of $j$,
so this generates all powers of $x$ simultaneously. This feature
is called {\it quantum parallelism}. Some of these powers are 1,
which correspond to the states $\sket 0 \sket 1$, $\sket r \sket
1$, $\sket {2r} \sket 1$, ..., $\sket{\left(\frac {2^t}{r}
-1\right)r} \sket 1$. This explains the choice (\ref{vx}) for
$V_x$. Classically, one would calculate successively $x^j$, for
$j$ starting from 2 until reaching $j=r$. \textit{Quantumly}, one
can calculate all powers of $x$ with just one application of
$V_x$. At the quantum level, the values of $j$ that yield $x^j
\equiv 1\mbox{ modulo }N$ are ``known''. But this quantum
information is not fully available at the classical level. A
classical information of a quantum state is obtained by practical
measurements and, at this point, it does not help if we measure
the first register, since all states in the superposition
(\ref{psi2}) have equal amplitudes. The first part of the strategy
to find $r$ is to observe that the first register of the states
$\sket 0 \sket 1$, $\sket r \sket 1$, $\sket {2r} \sket 1$, ...,
$\sket{2^t-r} \sket 1$ is periodic. So the information we want is
a period. In order to simplify the calculation, let us measure the
second register. Before doing this, we will rewrite $\sket{\j_2}$
collecting equal terms in the second register. Since $x^j$ is a
periodic function with period $r$, substitute $ar+b$ for $j$ in
Eq. (\ref{psi2}), where $0 \leq a \leq (2^t/r) - 1 $ and $0\leq b
\leq r-1$. Recall that we are supposing that $t=n$ and $r$ is a
power of 2, therefore $r$ divides $2^t$. Eq. (\ref{psi2}) is
converted to
\be
\sket{\j_2} = \frac{1}{\sqrt{2^t}} \sum_{b=0}^{r-1} \left(
\sum_{a=0}^{\frac{2^t}{r}-1}\sket{ar+b} \right) \sket {x^b}.
\label{psi2a} \ee
In the second register, we have substituted $x^b$ for $x^{ar+b}$,
since $x^r \equiv 1$ modulo $N$. Now the second register is
measured. Any output $x^0$, $x^1$, ..., $x^{r-1}$ can be obtained
with equal probability. Suppose that the result is $x^{b_0}$. The
state of the quantum computer is now
\be
\sket{\j_3} = \sqrt{\frac{r}{2^t}}
\left(\sum_{a=0}^{\frac{2^t}{r}-1} \sket{ar+b_0}\right) \sket
{x^{b_0}}. \label{psi3} \ee
Note that after the measurement, the constant is renormalized to
$\sqrt{r/2^t}$, since there are  $2^t/r$ terms in the sum
(\ref{psi3}).
Fig. \ref{graph} shows the probability of
obtaining the states of the computational basis upon measuring the
first register. The probabilities form a periodic function with
period $r$. Their values are zero except for the states
$\sket{b_0}$, $\sket{r+b_0}$, $\sket{2r+b_0}$, ...,
$\sket{2^t-r+b_0}$.

How can one find out the period of a function efficiently? The
answer is in the Fourier transform. The Fourier transform of a
periodic function with period $r$ is a new periodic function with
period proportional to $1/r$. This makes a difference for finding
$r$. The Fourier transform is the second and last part of the
strategy. The whole method relies on an efficient quantum
algorithm for calculating the Fourier transform, which is not
available classically. In Section \ref{universal}, we show that
the Fourier transform is calculated efficiently in a quantum
computer.

\section{The quantum discrete Fourier transform}

The Fourier transform of the function $F:\{0, \ldots, N-1\}
\rightarrow \mathbb{C}$ is a new function $\tilde F:\{0, \ldots,
N-1\} \rightarrow \mathbb{C}$ defined as
\be
\tilde F (k) = \frac{1}{\sqrt N} \sum_{j=0}^{N-1} e^{2 \p i j k/N}
F(j) \label{F}. \ee
We can apply the Fourier transform either on a function  or on the
states of the computational basis. The Fourier transform applied
to the state $\sket k$ of the computational basis $\{ \sket{0},
\dots, \sket{N-1} \}$ is
\be
\mbox{DFT}(\sket k) = \sket{\j_k} = \frac{1}{\sqrt N}
\sum_{j=0}^{N-1} e^{2 \p i j k/N} \sket j , \label{psia} \ee
where the set $\{ \sket{\j_k}: k=0, \ldots, N-1 \}$ forms a new
orthonormal basis. The Fourier transform is a unitary linear
operator. So, if we know how it acts on the states of the
computational basis, we also know how it acts on a generic state
\bestar \sket \j = \sum_{a=0}^{N-1} F(a) \sket a. \eestar
The Fourier transform of $ \sket \j$ can be performed indistinctly
using either (\ref{F}) or (\ref{psia}). We will use the latter.

\begin{figure}
    \setcaptionmargin{.5in}
    \centering
    \psfrag{2pi/7}{ $\frac{2\p}{7}$}
    \psfrag{Re}{\footnotesize Re}
    \psfrag{Im}{\footnotesize Im}
    \psfrag{j=0}[][]{\footnotesize$j=0$}
    \psfrag{j=1}[][]{\footnotesize$j=1$}
    \psfrag{j=2}[][]{\footnotesize$j=2$}
    \psfrag{j=3}[][]{\footnotesize$j=3$}
    \psfrag{j=4}[][]{\footnotesize$j=4$}
    \psfrag{j=5}[][]{\footnotesize$j=5$}
    \psfrag{j=6}[][]{\footnotesize$j=6$}
    \psfrag{j=7}[][]{\footnotesize$j=7$}
    \includegraphics[height=2.3in]{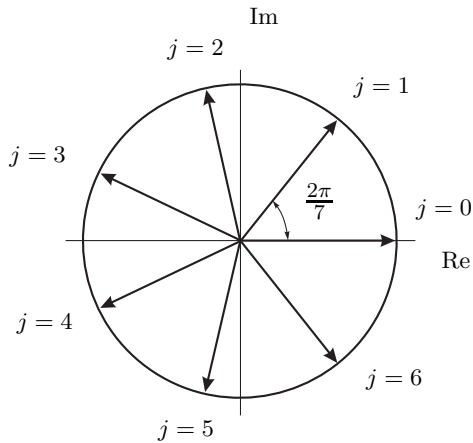}
    \caption[]{Vectors $e^{2 \p i j /7}$ $(j=0, ..., 6)$ in the
    complex plane. Their sum is zero by symmetry arguments. This
    is an example of Eq. (\ref{ident}) for $N=7$, $k=1$.}
    \label{2pi7}
\end{figure}

To prove that $\{ \sket{\j_k}: k=0, \ldots, N-1 \}$ is an
orthonormal basis, i.e.,
\bestar \sVEV{\j_{k^\prime}|\j_k} = \d_{k^\prime k}, \eestar
we can use the identity
\be
\frac{1}{N} \sum_{j=0}^{N-1} e^{2 \p i j k/N} = \left\{
\begin{array}{l} 1 \mbox{   if $k$ is a multiple of $N$} \\ 0
\mbox{   otherwise,} \end{array} \right. \label{ident} \ee
which is useful in the context of Fourier transforms. It is easy
to verify that (\ref{ident}) is true. If $k$ is a multiple of $N$,
then $e^{2 \p i j k/N} = 1$ and the first case of the identity
follows. If $k$ is not a multiple of $N$, (\ref{ident}) is true
even if $N$ is not a power of 2. Fig. \ref{2pi7} shows each term
$e^{2 \p i j k/N}$ $(j=0, ..., 6)$ for the case $k=1$ and $N=7$,
as vectors in the complex plane. Note that the sum of vectors must
be zero by a symmetry argument: the distribution of vectors is
isotropic. Usually it is said that the interference is destructive
in this case.
Using this identity, we can define the inverse Fourier transform,
which is similar to (\ref{psia}), just with a minus sign on the
exponent. Note that $\mbox{DFT}^{-1}= \mbox{DFT}^\dagger$, since
DFT is a unitary operator.

We will present the details of a quantum circuit to perform the
Fourier transform in Section \ref{universal}. Now we will continue
the calculation process of the circuit of Fig. \ref{order}. We are
ready to find out the next state of the quantum computer---$\sket
{\j_4}$. Applying the inverse Fourier transform on the first
register, using Eq. (\ref{psia}) and the linearity of
DFT$^\dagger$, we obtain
\beastar \sket{\j_4} & = & \mbox{DFT}^\dagger (\sket{\j_3}) \non
\\
 & = &
\sqrt{\frac{r}{2^t}} \sum_{a=0}^{\frac{2^t}{r}-1} \left (
\frac{1}{\sqrt {2^t}} \sum_{j=0}^{2^t-1} e^{-2 \p i j
(ar+b_0)/{2^t}} \sket j \right) \sket{x^{b_0}}. \label{psi4}
\eeastar
Inverting the summation order, we have
\be
\sket{\j_4} = \frac{1}{\sqrt r} \left(\sum_{j=0}^{2^t-1} \left[
\frac{1}{{2^t}/r} \sum_{a=0}^{\frac{2^t}{r}-1} e^\frac{-2 \p i j
a}{{2^t}/r} \right] e^{-2 \p i j b_0/2^t}
\sket{j}\right)\sket{x^{b_0}}. \label{psi4a} \ee
Using (\ref{ident}), we see that the expression in square brackets
is not zero if and only if $j= k{2^t}/r$, with $k = 0, ..., r-1 $.
When $j$ takes such values, the expression in the square brackets
is equal to 1. So we have
\be \sket{\j_4} = \frac{1}{\sqrt r} \left(\sum_{k=0}^{r-1} e^{-2
\p i \frac k r b_0} \sket{\frac{k 2^t}{r}}\right)\sket{x^{b_0}}.
\label{psi4b} \ee
In order to find $r$, the expression for $\sket{\j_4}$ has two
advantages over the expression for $\sket{\j_3}$ (Eq.
(\ref{psi3})): $r$ is in the denominator of the ket label and the
random parameter $b_0$ moved from the ket label to the exponent
occupying now a harmless place.
\begin{figure}
    \setcaptionmargin{.5in}
    \centering
    \psfrag{0}[][]{ \footnotesize $0$}
    \psfrag{b0}[][]{$\frac{2^t}{r}$}
    \psfrag{Rb0}[][]{$\frac{2\,2^t}{r}$}
    \psfrag{2rb0}[][]{$\frac{3\,2^t}{r}$}
    \psfrag{3rb0}[][]{$\frac{4\,2^t}{r}$}
    \psfrag{R2t}[][]{${\frac{1}{r}}$}
    \psfrag{R}[][]{$\frac{2^t}{r}$}
    \psfrag{Amplitude of each}{\footnotesize Probability distribution}
    \psfrag{terms}{\footnotesize Terms of $\sket{\psi_4}$}
    \psfrag{1st register}{\footnotesize (1st register)}
    \includegraphics[width=4.5in]{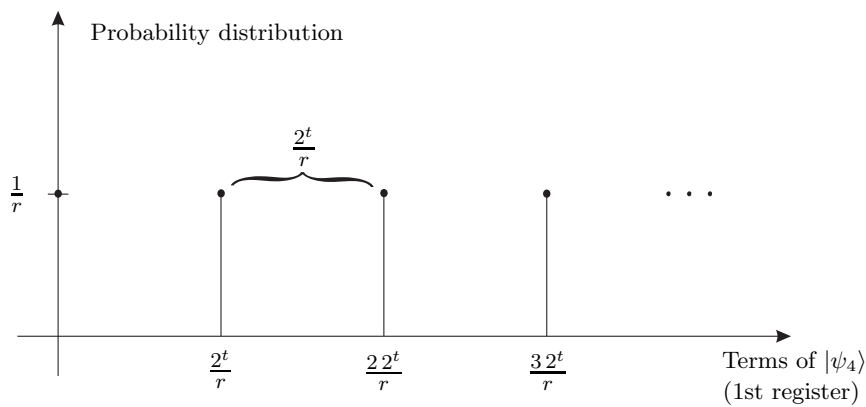}
    \caption[]{Probability distribution of $\sket {\psi_4}$
    measured in the computational basis. The horizontal axis
    has $2^t$ points, only the non-null terms are shown.
    The number of peaks is $r$ and the period is $2^t/r$.}
    \label{graph2}
\end{figure}
Fig. \ref{graph2} shows the probability distribution of
$\sket{\psi_4}$ measured in the computational basis. Measuring the
first register, we get the value $k_0 {2^t}/r$, where $k_0$ can be
any number between 0 and $r-1$ with equal probability (the peaks
in Fig. \ref{graph2}). If we obtain $k_0=0$, we have no clue at
all about $r$, and the algorithm must be run again. If $k_0 \neq
0$, we divide $k_0 {2^t}/r$ by $2^t$, obtaining $k_0/r$. Neither
$k_0$ nor $r$ are known. If $k_0$ is coprime to $r$, we simply
select the denominator.

If $k_0$ and $r$ have a common factor, the denominator of the
reduced fraction ${k_0}/r$ is a factor of $r$ but not $r$ itself.
Suppose that the denominator is $r_1$. Let $r= r_1 r_2$. Now the
goal is to find $r_2$, which is the order of $x^{r_1}$. We run
again the quantum part of the algorithm to find the order of
$x^{r_1}$. If we find $r_2$ in the first round, the algorithm
halts, otherwise we apply it recursively. The recursive process
does not last, because the number of iterations is less than or
equal to $\log_2 r$.

Take $N=15$ as an example, which is the least nontrivial composite
number. The set of numbers less than 15, coprime to 15 is
$\{1,2,4,7,8,11,13,14\}$. The numbers in the set $\{4,11,14\}$
have order 2 and the numbers in the set $\{2,7,8,13\}$ have order
4. Therefore, in any case $r$ is a power of 2 and the factors of
$N=15$ can be found in a 8-bit quantum computer ($t+n=2\lceil
\log_2 15 \rceil=8$). The authors of \cite{vandersypen} used a
7-qubit quantum computer, bypassing part of the algorithm.

\section{Generalization by means of an example}
\label{generalization}

In the previous sections, we have considered a special case when
the order $r$ is a power of 2 and $t=n$ ($t$ is the number of
qubits in the first register---see Fig. \ref{order}---and
$n=\lceil \log_2 N \rceil $). In this section, we consider the
factorization of $N=21$, that is the next nontrivial composite
number. We must choose $t$ such that $2^t$ is between $N^2$ and
$2N^2$, which is always possible \cite{shor}. For $N=21$, the
smallest value of $t$ is 9. This is the simplest example allowed
by the constraints, but enough to display all properties of Shor's
algorithm.

The first step is to pick up $x$ at random such that $1<x<N$, and
to test whether $x$ is coprime to $N$. If not, we easily find a
factor of $N$ by calculating GCD$(x,N)$. If yes, the quantum part
of the algorithm starts. Suppose that $x=2$ has been chosen. The
goal is to find out that the order of $x$ is $r=6$. The quantum
computer is initialized in the state
$$ \sket{\j_0} = \sket 0 \sket 0, $$
where the first register has $t=9$ qubits and the second has $n=5$
qubits. Next step is the application of $H^{\otimes 9}$ on the
first register yielding (see Eq. (\ref{psi1}))
$$ \sket {\j_1} = \frac{1}{\sqrt {512}} \sum_{j=0}^{511} \sket j
\sket 0 . $$
The next step is the application of $V_x$ (defined in (\ref{vx})),
which yields
\bea \sket {\j_2} & = & \frac{1}{\sqrt {512}} \sum_{j=0}^{511}
\sket j \sket{2^j \mbox{ mod }N } \non \\
 & = & \frac{1}{\sqrt{512}} \bigg( \;\; \sket{0}\sket{1}+\sket{1}\sket{2}+
\sket{2}\sket{4}+\sket{3}\sket{8}+ \;\; \sket{4}\sket{16}+ \;\,
\sket{5}\sket{11}+
 \non \\
 & & \;\;\;\;\;\;\;\;\;\;\;\;\;\;\, \sket{6}\sket{1}+\sket{7}\sket{2}+\sket{8}\sket{4}+\sket{9}\sket{8}+
\sket{10}\sket{16}+\sket{11}\sket{11}+ \non \\
 & & \;\;\;\;\;\;\;\;\;\;\;\;\, \sket{12}\sket{1}+ \ldots \bigg). \non \eea
Notice that the above expression has the following pattern: the
states of the second register of each ``column'' are the same.
Therefore we can rearrange the terms in order to collect the
second register:
\bea \sket {\j_2} & = & \frac{1}{\sqrt{512}} \bigg[ \big(\,
\sket{0}+\;\,\sket{6}+\sket{12}+ \ldots +\sket{504}+\sket{510}
\big) \, \sket{1}+ \non \\
 & & \;\;\;\;\;\;\;\;\;\;\;\, \big(
\sket{1}+\;\,\sket{7}+\sket{13}+ \ldots +\sket{505}+\sket{511}
\big) \sket{2}+ \non \\
 & & \;\;\;\;\;\;\;\;\;\;\;\, \big(
\sket{2}+\;\,\sket{8}+\sket{14}+ \ldots +\sket{506}\big)
\;\,\sket{4}+ \label{rearranged} \\
 & & \;\;\;\;\;\;\;\;\;\;\;\, \big(
\sket{3}+\;\,\sket{9}+\sket{15}+ \ldots +\sket{507}\big)
\;\,\sket{8}+ \non
\\
 & & \;\;\;\;\;\;\;\;\;\;\;\, \big(
\sket{4}+\sket{10}+\sket{16}+ \ldots +\sket{508}\big) \sket{16}+
\non
\\
 & & \;\;\;\;\;\;\;\;\;\;\; \big(
\sket{5}+\,\sket{11}+\sket{17}+ \ldots +\sket{509}\big) \sket{11}
\bigg]. \non \eea
This feature was made explicit in Eq. (\ref{psi2a}). Because the
order is not a power of 2, here there is a small difference: the
first two lines of Eq. (\ref{rearranged}) have 86 terms, while the
remaining ones have 85.

\begin{figure}
    \setcaptionmargin{.5in}
    \psfrag{Prob(j)}{\LARGE Prob($j$)}
    \psfrag{j}{\LARGE $j$}
    \hspace{0.4in}
    \includegraphics[angle=-90,totalheight=2.5in,width=4.5in]{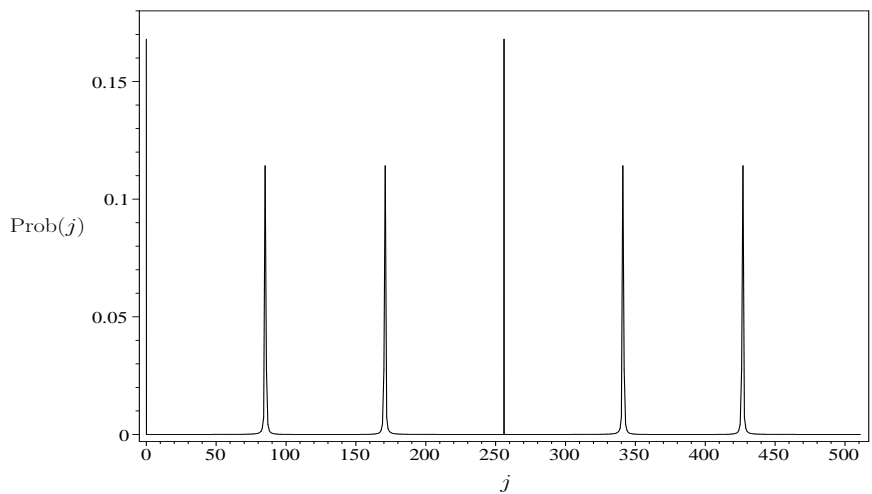}
    \caption{Plot of Prob$(j)$ against $j$. Compare to the plot
    of Fig. \ref{graph2}, where peaks are not spread and have the same height.}
    \label{probgraph}
\end{figure}

Now one measures the second register\footnote{As measurements can
always be performed in the end (see \cite{chuang} page 186), this
step is not necessary. It is commonly used to simplify the
expressions that follow.}, yielding one of the following numbers
equiprobably: $\{1,2,4,8,16,11\}$. Suppose that the result of the
measurement is 2, then
\be
\sket{\j_3} = \frac{1}{\sqrt{86}} \left(
\sket{1}+\sket{7}+\sket{13}+ \ldots +\sket{505}+\sket{511} \right)
\sket{2}. \label{psi3a} \ee
Notice that the state $\sket{\j_3}$ was renormalized in order to
have unit norm. It does not matter what is the result of the
measurement; what matters is the periodic pattern of
(\ref{psi3a}). The period of the states of the first register is
the solution to the problem and the Fourier transform can reveal
the value of the period. So, the next step is the application of
the inverse Fourier transform on the first register of
$\sket{\j_3}$:
\bea \sket{\j_4} & = & \mbox{DFT}^\dagger (\sket{\j_3}) \non \\
 & = & \mbox{DFT}^\dagger \left(
\frac{1}{\sqrt{86}} \sum_{a=0}^{85} \sket{6a+1} \right) \sket 2
 \non\\
 & = & \frac{1}{\sqrt{512}} \sum_{j=0}^{511} \left( \left[ \frac{1}{\sqrt{86}}
\sum_{a=0}^{85} e^{-2 \p i \frac{6ja}{512}} \right] e^{-2 \p i
\frac{j}{512}} \sket j\right) \sket{2}, \label{psi4ft} \eea
where we have used Eq. (\ref{psia}) and have rearranged the sums.
The last equation is similar to Eq. (\ref{psi4a}), but with an
important difference. In Section \ref{sectionorder}, we were
assuming that $r$ divides $2^t$. This is not true in the present
example (6 does not divide 512), therefore we cannot use the
identity (\ref{ident}) to simplify the term in brackets in Eq.
(\ref{psi4ft}). This term never vanishes, but its main
contribution is still around $j=0,\,85,\,171,$ $256,$ $341,$
$427,$ which are obtained rounding $512k_0/6$ for $k_0$ from 0 to
5---compare to the discussion that follows Eq. (\ref{psi4b}). To
see this, let us plot the probability of getting the result $j$
(in the interval 0 to 511) by measuring the first register of the
state $\sket {\psi_4}$. From (\ref{psi4ft}), we have that the
probability is
\be
\label{prob} \mbox{Prob}(j) = \frac{1}{512 \times 86} \left|
\sum_{a=0}^{85} e^{-2 \p i \frac{6ja}{512}} \right|^2. \ee
The plot of Prob$(j)$ is shown in Fig. \ref{probgraph}.
We see the peaks around $j=0,\,85,\,171,$ $256,$ $341,$ $427,$
indicating a high probability of getting one of these values, or
some value very close to them. In between, the probability is
almost zero. The sharpness of the peaks depends on $t$ (number of
qubits in the first register). The lower limit $2^t \geq N^2$
ensures a high probability in measuring a value of $j$ carrying
the desired information. A careful analysis of the expression
(\ref{prob}) is performed in \cite{lomonaco} and a meticulous
study of the peak form is performed in \cite{einarsson}.

Let us analyze the possible measurement results. If we get $j=0$
(first peak), the algorithm has failed in this round. It must be
run again. We keep $x=2$ and rerun the quantum part of the
algorithm. The probability of getting $j=0$ is low: from Eq.
(\ref{prob}) we have that Prob$(0) = 86/512 \approx 0.167$. Now
suppose we get $j=85$ (or any value in the second peak). We divide
by 512 yielding $85/512$, which is a rational approximation of
$k_0/6$, for $k_0=1$. How can we obtain $r$ from $85/512$?

The method of continued fraction approximation allows one to
extract the desired information. A general continued fraction
expansion of a rational number $j_1/j_2$ has the form
$$ \frac{j_1}{j_2} = a_0 +
\frac{1}{a_1+\frac{1}{\ldots+\frac{1}{a_p}}}, $$
usually represented as $[a_0,a_1,...,a_p]$, where $a_0$ is a
non-negative integer and $a_1,...,a_p$ are positive integers. The
$q$-th convergent ($0\leq q \leq p$) is defined as the rational
number $[a_0,a_1,...,a_q]$. It is an approximation to $j_1/j_2$
and has a denominator smaller than $j_2$.

This method is easily applied by inversion of the fraction
followed by integer division with rational remainder. Inverting
$85/512$ yields $512/85$, which is equal to $6+2/85$. We repeat
the process with $2/85$ until we get numerator 1. The result is
\bestar \frac{85}{512} = \frac{1}{6+\frac{1}{42+\frac{1}{2}}}.
\eestar
So, the convergents of $85/512$ are $1/6$, $42/253$, and $85/512$.
We must select the convergents that have a denominator smaller
than $N=21$ (since $r < N$)\footnote{The inequality  $r \leq
\varphi(N)$ follows from the Euler's theorem:
$x^{\varphi(N)}\equiv1 \mod N$, where $x$ is a positive integer
coprime to $N$ and $\varphi$ is the Euler's totient function
($\varphi(N)$ gives the number of positive integers less than $N$,
coprime to $N$). The inequality $\varphi(N)<N$ follows from the
definition of $\varphi$. (see \cite{von} page 492) }. This method
yields $1/6$, and then $r=6$. We check that $2^6 \equiv 1$ modulo
21, and the quantum part of the algorithm ends with the correct
answer. The order $r=6$ is an even number, therefore
GCD$(2^{(6/2)} \pm 1, 21)$ gives two non trivial factors of $21$.
A straightforward calculation shows that any measured result in
the second peak (say $81\leq j \leq 89$) yields the convergent
1/6.

Consider now the third peak, which corresponds to $k_0/6$,
$k_0=2$. We apply again the method of continued fraction
approximation, which yields $1/3$, for any $j$ in the third peak
(say $167\leq j \leq 175$). In this case, we have obtained a
factor of $r$ $(r_1=3)$, since $2^3 \equiv 8 \not\equiv 1$ modulo
21. We run the quantum part of the algorithm again to find the
order of 8. We eventually obtain $r_2=2$, which yields $r=r_1 r_2
= 3\times 2=6$.

The fourth and fifth peaks yield also factors of $r$. The last
peak is similar to the second, yielding $r$ directly.

The general account of the succeeding probability is as follows.
The area under all peaks is approximately the same:
$\approx0.167$. The first and fourth peaks have a nature different
from the others---they are not spread. To calculate their
contribution to the total probability, we take the basis equal to
1. The area under the second, third, fifth, and last peaks are
calculated by adding up Prob$(j)$, for $j$ running around the
center of each peak. So, in approximately 17\% cases, the
algorithm fails (1st peak). In approximately 33\% cases, the
algorithm returns $r$ in the first round (2nd and 6th peaks). In
approximately 50\% cases, the algorithm returns $r$ in the second
round or more (3rd, 4th, and 5th peaks). Now we calculate the
probability of finding $r$ in the second round. For the 3rd and
5th peaks, the remaining factor is $r_2=2$. The graph equivalent
to Fig. \ref{probgraph} in this case has 2 peaks, then the
algorithm returns $r_2$ in 50\% cases. For the 4th peak, the
remaining factor is $r=3$ and the algorithm returns $r_2$ in
66.6\% cases. This amounts to $\frac{2\times50\%+66.6\%}{3}$ of
50\%, which is equal to  around 22\%. In summary, the success
probability for $x=2$ is around 55\%.

\section{Fourier transform in terms of the universal gates}
\label{universal}

In the previous section, we have shown that Shor's algorithm is an
efficient probabilistic algorithm, assuming that the Fourier
transform could be implemented efficiently. In this section, we
decompose the Fourier transform in terms of the universal gates:
CNOT and 1-qubit gates. This decomposition allows one to measure
the efficiency of the quantum discrete Fourier transform and shows
how to implement it in an actual quantum computer.

The Fourier transform of the states of the computational basis is
\be \mbox{DFT}(\sket j) = \frac{1}{\sqrt N} \sum_{k=0}^{N-1} e^{2
\p i j k/N} \sket k . \label{FT}\ee
Noting that the right hand side of Eq. (\ref{FT}) has $N$ terms
and the computational basis has $N$ states, we derive that the
complexity to calculate classically the Fourier transform of the
computational basis using Eq. (\ref{FT}) is $O(N^2) = O(2^{2n})$
-- double exponential growth. A very important result in Computer
Science was the development of the classical fast Fourier
transform (FFT), which reduced the complexity to $O(n2^n)$
\cite{cooley}. In the present context we show the improvement by
recognizing that the rhs of (\ref{FT}) is a very special kind of
expansion, which can be fully factored. For example, the Fourier
transform of \{$\sket{0}$, $\sket{1}$, $\sket{2}$, $\sket{3}$\}
can be written as
\bea
\mbox{DFT}(\sket 0) & = & \left( \frac{\sket 0 + \sket 1}{\sqrt 2}
\right) \otimes \left( \frac{\sket 0 + \sket 1}{\sqrt 2} \right)
\non \\
\mbox{DFT}(\sket 1) & = & \left( \frac{\sket 0 - \sket 1}{\sqrt 2}
\right) \otimes \left( \frac{\sket 0 + i \sket 1}{\sqrt 2} \right)
\non \\
\mbox{DFT}(\sket 2) & = & \left( \frac{\sket 0 + \sket 1}{\sqrt 2}
\right) \otimes \left( \frac{\sket 0 - \sket 1}{\sqrt 2} \right)
\label{example} \\
\mbox{DFT}(\sket 3) & = & \left( \frac{\sket 0 - \sket 1}{\sqrt 2}
\right) \otimes \left( \frac{\sket 0 - i \sket 1}{\sqrt 2}
\right). \non
\eea
Note that in example (\ref{example}), we are using base 2 in order
to factor the rhs. Let us now factor the general expression. The
first step is to write (\ref{FT}) in the form
\be
\mbox{DFT}(\sket j) = \frac{1}{\sqrt {2^{n}}} \sum_{k_1=0}^{1}
\ldots \sum_{k_n=0}^{1} e^{2 \p i j \sum_{l=1}^n \frac{k_l}{2^l}}
\sket{k_1} \otimes \ldots \otimes \sket{k_n}, \label{DFTj}
\ee
where the ket $\sket k$ was converted to base 2 and we have used
the expansion $k = \sum_{l=1}^n k_l 2^{n-l}$ in the exponent.
Using that the exponential of a sum is a product of exponentials,
(\ref{DFTj}) turns into a (non-commutative) product of the
following kets:
\be
\label {ft2} \mbox{DFT}(\sket j) = \frac{1}{\sqrt {2^{n}}}
\sum_{k_1=0}^{1} \ldots \sum_{k_n=0}^{1} \prod_{l=1}^n \left( e^{2
\p i j \frac{k_l}{2^l}} \sket{k_l} \right).
\ee
Now we factor (\ref{ft2}) by interchanging the sums and the
product:
\be
\label{ft3} \mbox{DFT}(\sket j) = \frac{1}{\sqrt {2^{n}}}
\prod_{l=1}^n \sum_{k_l=0}^{1} \left( e^{2 \p i j \frac{k_l}{2^l}}
\sket{k_l} \right).
\ee
We easily convince ourselves that the last equation is correct by
going backwards: simply expand the product in Eq. (\ref{ft3}) and
then put all sums at the beginning of the resulting expression to
obtain (\ref{ft2}). Expanding the sum of Eq. (\ref{ft3}) and then
the product, we finally get
\bea \mbox{DFT}(\sket j) & = &
\frac{1}{\sqrt{2^{n}}} \prod_{l=1}^n \left( \sket 0 + e^{2 \p i j
/ 2^l} \sket 1 \right) \non \\
 & = & \!\!\!\! \left( \frac{\sket 0 + e^{2 \p i \frac{j}{2} } \sket 1}{\sqrt 2}
 \right)\! \otimes \!\left( \frac{\sket 0 + e^{2 \p i \frac{j}{2^2}} \sket 1}{\sqrt 2}
 \right) \!\otimes\! \ldots \!\otimes \!\left( \frac{\sket 0 + e^{2 \p i \frac{j}{2^n}} \sket 1}{\sqrt 2}
 \right). \label{ftj}
\eea The complexity to calculate Eq. (\ref{ftj}) for one $\sket j$
is $O(n)$, since there are $n$ terms in the product. The
complexity in the classical calculation of the fast Fourier
transform of the whole computational basis is still exponential --
$O(n2^n)$, since the calculation is performed on each of the $2^n$
basis elements, one at a time. On the other hand, the quantum
computer uses quantum parallelism, and the Fourier transform of
the state
$$ \sket \j = \sum_{a=0}^{2^n-1} F(a) \sket a, $$
that has an exponential number of terms, is calculated with one
application of the quantum Fourier transform. The Fourier
transform of the $2^n$ basis elements is performed simultaneously,
so the complexity of the quantum Fourier transform is measured by
the size of its circuit. We now show that it requires $O(n^2)$
gates.
\begin{figure}
    \setcaptionmargin{.5in}
    \centering
    \psfrag{R2}{\footnotesize$R_2$}
    \psfrag{Rn+1-l}[][]{\footnotesize$R_{n+1-l}$}
    \psfrag{H}{\footnotesize$H$}
    \psfrag{j1}[][]{\footnotesize$\sket{j_1}$}
    \psfrag{jl-1}[][]{\footnotesize$\sket{j_{l-1}}$}
    \psfrag{jl}[][]{\footnotesize$\sket{j_{l}}$}
    \psfrag{jl+1}[][]{\footnotesize$\sket{j_{l+1}}$}
    \psfrag{jn}[][]{\footnotesize$\sket{j_{n}}$}
    \psfrag{psi}[][]{\footnotesize$\sket{\j}$}
    \includegraphics[]{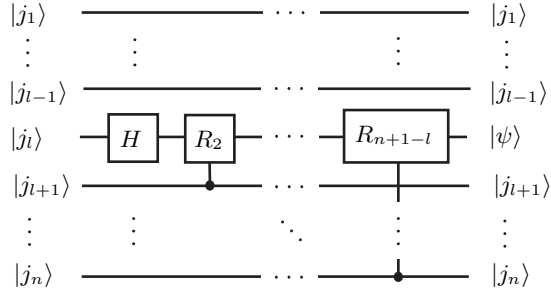}
    \caption[]{Part of the quantum Fourier transform
circuit that acts on qubit $\sket{j_l}$. The value of all qubits
does not change, except $\sket{j_l}$ that changes to $\sket{\j} =
\frac{\sket 0 + e^{2 \p i \frac{j}{2^{n+1-l}}} \sket 1}{\sqrt
2}.$}\label{FT(j)}
\end{figure}

Consider the circuit of Fig. \ref{FT(j)}. It is easy to check that
the value of the qubits $\sket{j_m}$, $m \neq l$, does not change.
Let us now check the hard one: $\sket{j_l}$. The unitary matrices
$R_k$ are defined as
\bestar R_k = \left[ \begin{array}{cc} 1 & 0 \\ 0 & \exp\left(2 \p
i \frac{1}{2^k}\right) \end{array} \right].  \eestar
Each $R_k$ gate is controlled by the qubit $\sket{j_{k+l-1}}$. So,
if $j_{k+l-1}=0$, then $R_k$ must be replaced by the identity
matrix (no action), and if $j_{k+l-1}=1$, then $R_k$ comes in
action. This means that, for calculation purposes, the $R_k$'s
controlled by $\sket{j_{k+l-1}}$ can be replaced by the 1-qubit
gates
\be
{\rm\textit{CR}}_k = \left[ \begin{array}{cc} 1 & 0 \\ 0 & \exp
\left( 2 \p i \frac{j_{k+l-1}}{2^k} \right) \end{array} \right].
\label{CRk}\ee
In order to simplify the calculations, note that
\be
H \sket{j_l} = \frac{\sket 0 + e^{2 \p i \frac{j_l}{2}}
\sket1}{\sqrt 2} = {\textit CR}_1 \sket + , \label{Hj} \ee
where $\sket + = \frac{1}{\sqrt 2} (\sket 0 +\sket 1 )$. So
instead of using
\bestar \sket \j = {\textit CR}_{n+1-l} \ldots {\textit CR}_2 \; H
\sket{j_l}, \eestar
which can be read directly from Fig. \ref{FT(j)}, we will use
\bestar \sket \j = {\textit CR}_{n+1-l} \ldots {\textit CR}_2
{\textit CR}_1 \sket{+}. \eestar
We define
\be \label{pr} {\rm \textit{PR}}_{n+1-l} = \prod_{k=n+1-l}^{1}
{\textit CR}_k, \ee
\begin{figure}
    \setcaptionmargin{.5in}
    \centering
    \psfrag{PR1}[][]{\footnotesize\textit{PR}$_1$}
    \psfrag{PR2}[][]{\footnotesize\textit{PR}$_2$}
    \psfrag{PRn-1}[][]{\footnotesize\textit{PR}$_{n-1}$}
    \psfrag{PRn}[][]{\footnotesize\textit{PR}$_n$}
    \psfrag{j1}[][]{\footnotesize$\sket{+}$}
    \psfrag{j2}[][]{\footnotesize$\sket{+}$}
    \psfrag{j3}[][]{\footnotesize$\sket{+}$}
    \psfrag{jn-1}[][]{\footnotesize\!\!\!$\sket{+}$}
    \psfrag{jn}[][]{\footnotesize$\sket{+}$}
    \psfrag{n}[][]{\scriptsize{\mbox{$\frac{\sket 0 + e^{2 \p i \frac{j}{2^n}} \sket 1}{\sqrt
    2}$}}}
    \psfrag{n-1}[][]{\scriptsize{\mbox{$\frac{\sket 0 + e^{2 \p i \frac{j}{2^{n-1}}} \sket 1}{\sqrt
    2}$}}}
    \psfrag{2}[][]{\scriptsize{\mbox{$\frac{\sket 0 + e^{2 \p i \frac{j}{2^2}} \sket 1}{\sqrt
    2}$}}}
    \psfrag{1}[][]{\scriptsize{\mbox{$\frac{\sket 0 + e^{2 \p i \frac{j}{2}} \sket 1}{\sqrt 2}$}}}
    \includegraphics[]{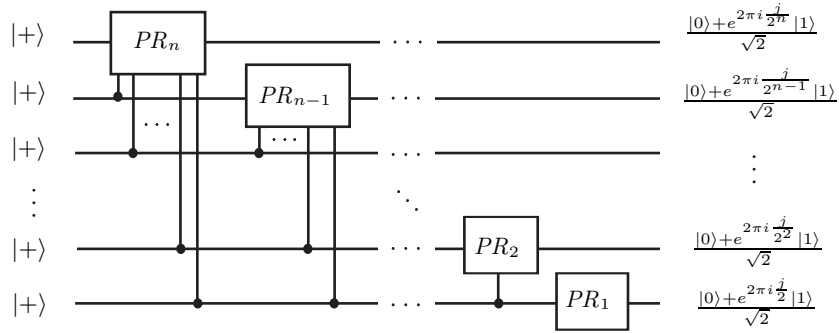}
    \renewcommand{\figurename}{Figure}
    \caption[]{Intermediate circuit for the quantum Fourier
Transform. The input is taken as $\sket +$ for calculation
purposes as explained in Eq. (\ref{Hj}). The output is in reverse
order with respect to Eq. (\ref{ftj}).}
    \label{intermediate}
\end{figure}
where the product is in the reverse order. Using (\ref{CRk}) and
(\ref{pr}), we get
\bea \label{Prn+1-l} {\rm \textit{PR}}_{n+1-l} & = &
\frac{1}{\sqrt{2}} \left[
\begin{array}{cc} 1 & 0
\\ 0 & \mbox{{$\exp  2 \p i \mbox{{$
\left(\frac{j_n}{2^{n+1-l}}+ \ldots + \frac{j_l}{2} \right)$}} $}}
\end{array}
\right] \non \\
 & = & \frac{1}{\sqrt{2}} \left[
\begin{array}{cc} 1 & 0
\\ 0 & \exp {\left( 2 \p i \frac{j}{2^{n+1-l}} \right) } \label{matrix} \end{array} \right],
 \eea
where we have used that $j = \sum_{m=1}^{n} j_m 2^{n-m}$ and the
fact that the first $l-1$ terms of this expansion do not
contribute---they are integer multiples of $2 \p i $ in
(\ref{matrix}). We finally get
\bea \sket \j & = & {\rm \textit{PR}}_{n+1-l} \sket + \non \\ & =
& \frac{\sket 0 + e^{2 \p i \frac{j}{2^{n+1-l}}} \sket 1}{\sqrt
2}. \eea
Note that ${\rm \textit{PR}}_{n+1-l}$ cannot be implemented
directly acting only in the $l$-th qubit because it requires the
values of $j_{l+1}$ to $j_n$.

\begin{figure}
    \setcaptionmargin{.5in}
    \centering
    \psfrag{=}[][]{$=$}
    \psfrag{phi}[][]{\footnotesize$\sket{\varphi}$}
    \psfrag{psi}[][]{\footnotesize$\sket{\j}$}
    \includegraphics[]{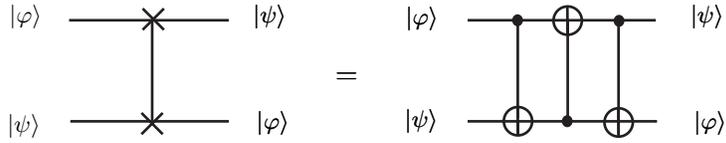}
    \renewcommand{\figurename}{Figure}
    \caption[]{The swap circuit.}
    \label{swap}
\end{figure}

The next step is the circuit of Fig. \ref{intermediate}. We have
merged the $R_k$ gates using Eq. (\ref{pr}). The gates \textit
{PR}$_k$ ($k$ from $n$ to 1) are placed in sequence in Fig.
\ref{intermediate}, so that the output of the first qubit is the
last term of Eq. (\ref{ftj}), corresponding to the action of
\textit{PR}$_n$ on $\sket{\j_1}$ controlled by the other qubits,
which do not change. The same process is repeated by \textit{
PR}$_{n-1}$ acting on $\sket{\j_2}$, yielding the term before the
last in Eq. (\ref{ftj}), and so on, until reproducing all the
terms of the Fourier transform. Now it remains to reverse the
order of the states of the qubits.

\begin{figure}
    \setcaptionmargin{.5in}
    \centering
    \psfrag{=}[][]{$=$}
    \psfrag{Rk}[][]{\footnotesize$R_k$}
    \psfrag{Rk-1}[][]{\footnotesize$R_k$}
    \psfrag{Rk-1+}[][]{\footnotesize$R_k^\dagger$}
    \includegraphics[]{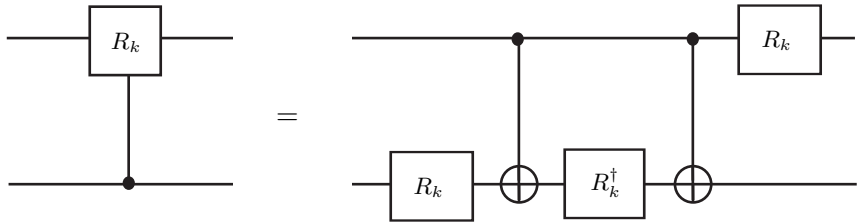}
    \renewcommand{\figurename}{Figure}
    \caption[]{Decomposition of the controlled $R_k$ gates in
    terms of the universal gates.}
    \label{decomposition}
\end{figure}

In order to reverse the states of 2 generic qubits, we use the
circuit of Fig. \ref{swap}. Let us show why this circuit works as
desired. Take the input $\sket \varphi\sket \psi=\sket 0\sket 1$.
The first CNOT of Fig. \ref{swap} does not change this state; the
upside down CNOT changes to $\sket 1\sket 1$; and the last CNOT
changes to $\sket 1\sket 0$. The output is $\sket \psi\sket
\varphi$. If we repeat the same process with $\sket 0\sket 0$,
$\sket 1\sket 0$, and $\sket 1\sket 1$, we conclude that the
circuit inverts all states of the computational basis, therefore
it inverts a generic state of the form $\sket \varphi\sket \psi$.

\begin{figure}
    \setcaptionmargin{.5in}
    \centering
    \psfrag{FT with}[][]{\small DFT with}
    \psfrag{reverse}[][]{\small reverse}
    \psfrag{output}[][]{\small output}
    \psfrag{j1}[][]{\footnotesize$\sket{j_1}$}
    \psfrag{j2}[][]{\footnotesize$\sket{j_2}$}
    \psfrag{jn-1}[][]{\footnotesize$\sket{j_{n-1}}$}
    \psfrag{jn}[][]{\footnotesize$\sket{j_{n}}$}
    \psfrag{n}[][]{\scriptsize{\mbox{$\frac{\sket 0 + e^{2 \p i \frac{j}{2^n}} \sket 1}{\sqrt
    2}$}}}
    \psfrag{n-1}[][]{\scriptsize{\mbox{$\frac{\sket 0 + e^{2 \p i \frac{j}{2^{n-1}}} \sket 1}{\sqrt
    2}$}}}
    \psfrag{2}[][]{\scriptsize{\mbox{$\frac{\sket 0 + e^{2 \p i \frac{j}{2^2}} \sket 1}{\sqrt
    2}$}}}
    \psfrag{1}[][]{\scriptsize{\mbox{$\frac{\sket 0 + e^{2 \p i \frac{j}{2}} \sket 1}{\sqrt 2}$}}}
    \includegraphics[]{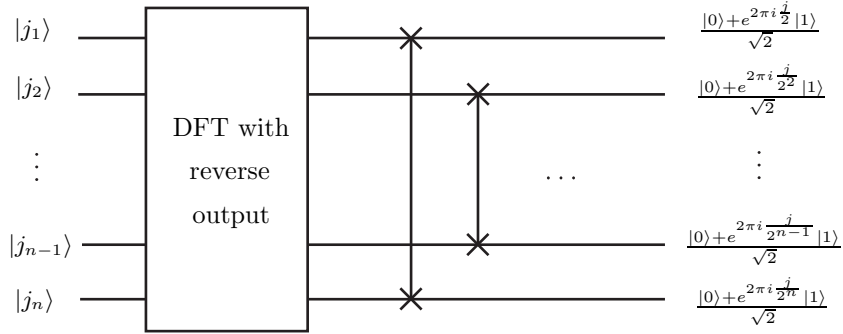}
    \renewcommand{\figurename}{Figure}
    \caption[]{The complete circuit for the quantum Fourier
Transform.}
    \label{complete}
\end{figure}

The decomposition is still not complete. It remains to write the
controlled $R_k$ gates in terms of CNOT and 1-qubit gates. This
decomposition is given in Fig. \ref{decomposition}. The
verification of this decomposition is straightforward. One simply
follows what happens to the computational basis $\{ \sket{00},
\sket{01}, \sket{10}, \sket{11} \} $ in both circuits.

The complete circuit for the quantum Fourier transform is given in
Fig. \ref{complete}. Now we can calculate the complexity of the
quantum Fourier circuit. Counting the number of elementary gates
in Figs. \ref{FT(j)} to \ref{decomposition} we get the leading
term $5n^2/2$, which implies that the complexity is $O(n^2)$.

By now one should be asking about the decomposition of $V_x$ in
terms of the elementary gates. $V_x$ is the largest gate of Fig.
\ref{order}. Actually, Shor stated in his 1997 paper that $V_x$ is
the ``bottleneck of the quantum factoring algorithm'' due to the
time and space consumed to perform the modular exponentiation (see
\cite{shor} page 10). The bottleneck is not so strict though
since, by using the well known classical method of repeated
squaring and ordinary multiplication algorithms (see \cite{von}
page 69), the complexity to calculate modular exponentiation is
$O(n^3)$. The quantum circuit can be obtained from the classical
circuit by replacing the irreversible classical gates by the
reversible quantum counterpart. $V_x$ is a problem in recursive
calls of the algorithm when $x$ changes. For each $x$, a new
circuit must be built, what is troublesome at the present stage of
hardware development.

%
%

\section*{Acknowledgments}

We thank the Group of Quantum Computation at LNCC for stimulating
discussions on the subject.


\end{document}